\documentclass[twocolumn]{aastex63}
\usepackage{ulem} 

\begin{document}

\title{FAST discovery of a fast neutral hydrogen outflow}

\author{Renzhi Su}
\email{surenzhi@zhejianglab.com}
\affiliation{Research Center for Intelligent Computing Platforms, Zhejiang Laboratory, Hangzhou 311100, China}
\affiliation{Key Laboratory for Research in Galaxies and Cosmology, Shanghai Astronomical Observatory, Chinese Academy of Sciences, \\80 Nandan 
Road, Shanghai 200030, China}
\affiliation{University of Chinese Academy of Sciences, 19A Yuquan Road, Beijing 100049, China \\}
\affiliation{ATNF, CSIRO Space and Astronomy, PO Box 76, Epping, NSW 1710, Australia}

\author{Minfeng Gu}
\email{gumf@shao.ac.cn}
\affiliation{Key Laboratory for Research in Galaxies and Cosmology, Shanghai Astronomical Observatory, Chinese Academy of Sciences, \\80 Nandan 
Road, Shanghai 200030, China}

\author{S. J. Curran}
\affiliation{School of Chemical and Physical Sciences, Victoria University of Wellington, PO Box 600, Wellington 6140, New Zealand}

\author{Elizabeth K. Mahony}
\affiliation{ATNF, CSIRO Space and Astronomy, PO Box 76, Epping, NSW 1710, Australia}

\author{Ningyu Tang}
\affiliation{Department of Physics, Anhui Normal University, Wuhu, Anhui 241002, People's Republic of China}

\author{James R. Allison}
\affiliation{First Light Fusion Ltd., Unit 9/10 Oxford Pioneer Park, Mead Road, Yarnton, Kidlington OX5 1QU, UK}

\author{Di Li}
\affiliation{National Astronomical Observatories, Chinese Academy of Sciences, 20A Datun Road, Beijing 100101, China}
\affiliation{Research Center for Intelligent Computing Platforms, Zhejiang Laboratory, Hangzhou 311100, China}
\affiliation{NAOC–UKZN Computational Astrophysics Centre, University of KwaZulu-Natal, Durban 4000, South Africa}

\author{Ming Zhu}
\affiliation{National Astronomical Observatories, Chinese Academy of Sciences, 20A Datun Road, Beijing 100101, China}
\affiliation{CAS Key Laboratory of FAST, National Astronomical Observatories, Chinese Academy of Sciences, Beijing 100101, China}

\author{J. N. H. S. Aditya}
\affiliation{Sydney Institute for Astronomy, School of Physics A28, University of Sydney, NSW 2006, Australia}
\affiliation{ARC Centre of Excellence for All Sky Astrophysics in 3 Dimensions (ASTRO 3D)}

\author{Hyein Yoon}
\affiliation{Sydney Institute for Astronomy, School of Physics A28, University of Sydney, NSW 2006, Australia}
\affiliation{ARC Centre of Excellence for All Sky Astrophysics in 3 Dimensions (ASTRO 3D)}

\author{Zheng Zheng}
\affiliation{National Astronomical Observatories, Chinese Academy of Sciences, 20A Datun Road, Beijing 100101, China}

\author{Zhongzu Wu}
\affiliation{College of Physics, Guizhou University, 550025 Guiyang, PR China}



\begin{abstract}
In this letter, we report the discovery of a fast neutral hydrogen outflow in SDSS J145239.38+062738.0, a merging radio galaxy containing an optical type I active galactic nuclei (AGN). This discovery was made through observations conducted by the Five-hundred-meter Aperture Spherical radio Telescope (FAST) using redshifted 21-cm absorption. The outflow exhibits a blueshifted velocity likely up to $\sim-1000\,\rm km\,s^{-1}$ with respect to the systemic velocity of the host galaxy with an absorption strength of $\sim -0.6\,\rm mJy\,beam^{-1}$ corresponding to an optical depth of 0.002 at $v=-500\,\rm km\,s^{-1}$. The mass outflow rate ranges between $2.8\times10^{-2}$ and $3.6\, \rm M_\odot \, yr^{-1}$, implying an energy outflow rate ranging between $4.2\times10^{39}$ and $9.7\times10^{40}\rm\,erg\,s^{-1}$, assuming 100 K $<T_{\rm s}<$ 1000 K. Plausible drivers of the outflow include the star bursts, the AGN radiation, and the radio jet, the last of which is considered the most likely culprit according to the kinematics. By analysing the properties of the outflow, the AGN, and the jet, we find that if the HI outflow is driven by the AGN radiation, the AGN radiation seems not powerful enough to provide negative feedback whereas the radio jet shows the potential to provide negative feedback. Our observations contribute another example of a fast outflow detected in neutral hydrogen, as well as demonstrate the capability of FAST in detecting such outflows.

\end{abstract}


\section{Introduction}\label{sec:intro}
\subsection{Background}
Redshifted HI 21-cm absorption provides us with unique insights into the neutral gas in and around galaxies throughout the Universe \citep[e.g.][]{morganti2015cool,allison2022the}. Depending on whether the absorbing neutral hydrogen gas is along the line-of-sight or within the host galaxy of the background radio continuum source, HI absorption is classified as intervening \citep[e.g.][]{kanekar2009a,kanekar2014the,curran2021inter,gupta2021blind,mahony2022hi} or associated \citep[e.g.][]{allison2012a,allison2014a,curran2016a,maccagni2017kinem,gupta2021evolu,murthy2021the,su2022flash}.

While intervening HI absorbers are good tracers of cold neutral gas reservoir for star formation \citep[e.g.][]{curran2017the,curran2019the}, associated HI absorbers have been frequently utilized to investigate the accretion and feedback processes in active galactic nuclei (AGN)\citep[e.g.][]{morganti2013radio,mahony2013the,maccagni2014what,aditya2019ugmrt,su2023does}. Generally, symmetric HI absorption components with full width at half maximum (FWHM) from several tens to a few hundreds $\rm km\,s^{-1}$ centred on or close to the systemic velocity trace regularly rotating structures whereas asymmetric, broader, shallow, and blueshifted HI absorption components indicate the presence of outflows \citep[e.g.][]{morganti2005fast,curran2016a}. Outflows play an important role in galaxy evolution \citep[e.g.][]{fabian2012obser,veilleux2020cool}. However, currently direct detections of such HI outflows are quite rare, see the compilation by \cite{morganti2018the}.

The detection of an outflow required a relatively large bandwidth in addition to sufficient sensitivity to detect low optical depths $\tau\sim0.005$\citep{morganti2018the}. If the background radio emission is weak (e.g. $<$ 500 mJy), the detection becomes unfeasible over a reasonable observing time. The advent of the Five-hundred-meter Aperture Spherical radio Telescope \citep[FAST;][]{nan2011the,nan2016fast,li2018fast,jiang2019commi,jiang2020the} gives us an ideal tool to explore such shallow HI outflows by utilizing its large collecting area. Recent observations have demonstrated the FAST capability in detecting HI absorption lines \citep{zhang2021extra,hu2023detec}.   
    
\subsection{Optical and radio properties of SDSS J145239.38+062738.0}   
\begin{figure}
	\includegraphics[width=8cm]{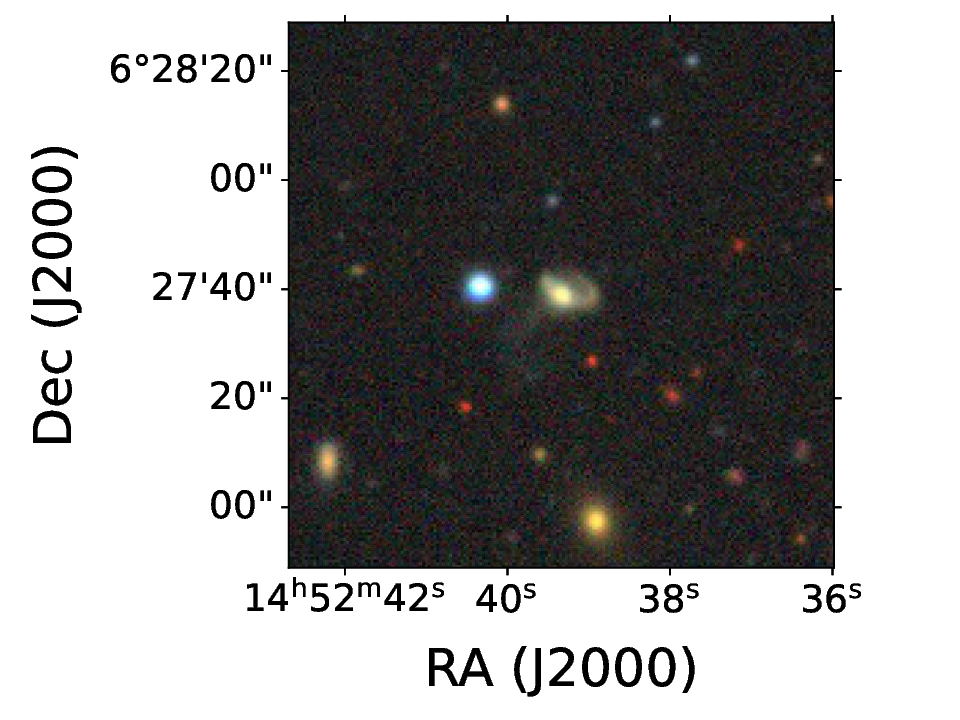}
    \caption{The $grz$ composite image of J1452+0627 (the central object) from the DESI Legacy Imaging Surveys \citep{dey2019the}, showing compelling evidence of tidal features indicating that J1452+0627 is a merging system. }
    \label{fig:J1426+0627_opt}
\end{figure}

\begin{figure*}
	\includegraphics[height=7cm]{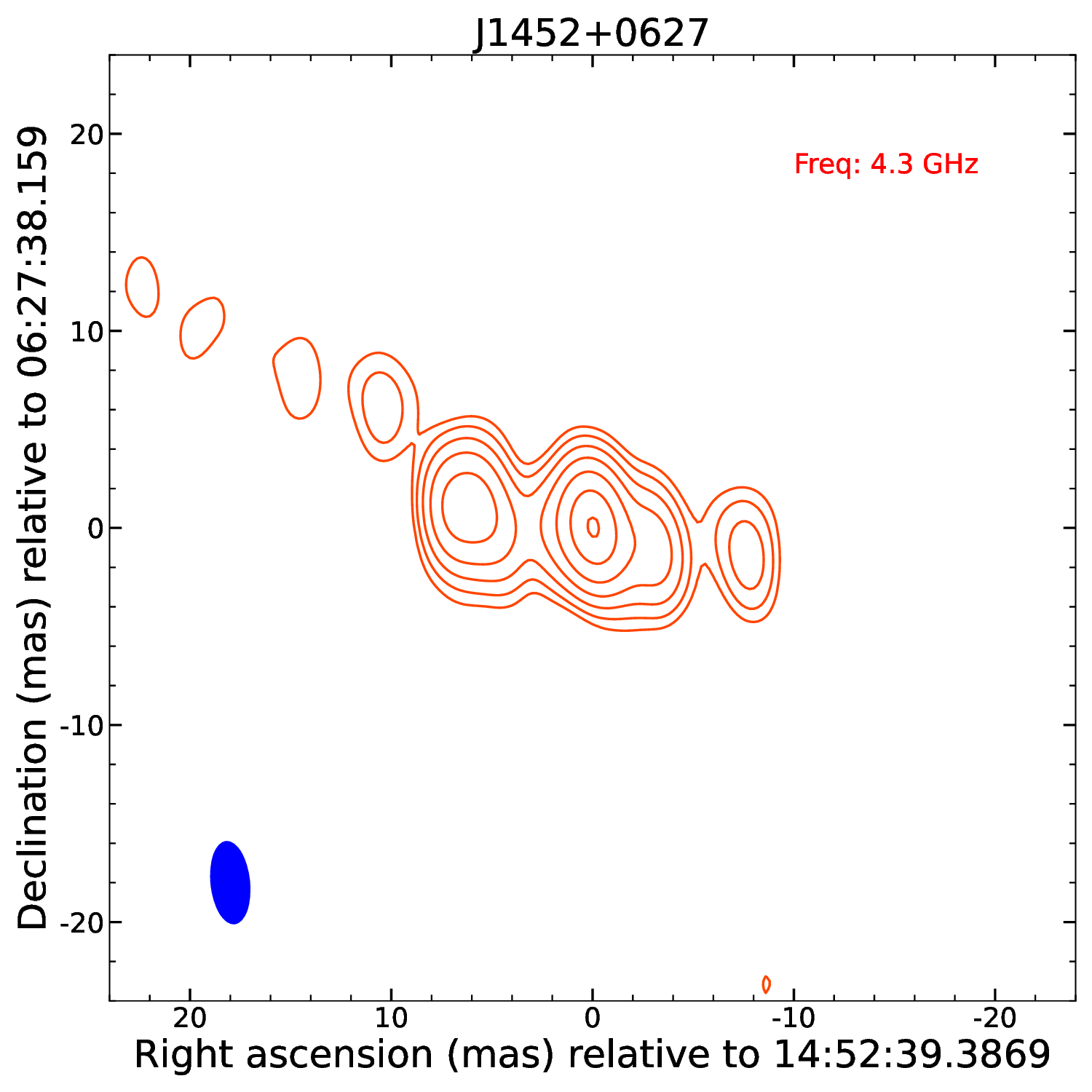}\includegraphics[height=7cm]{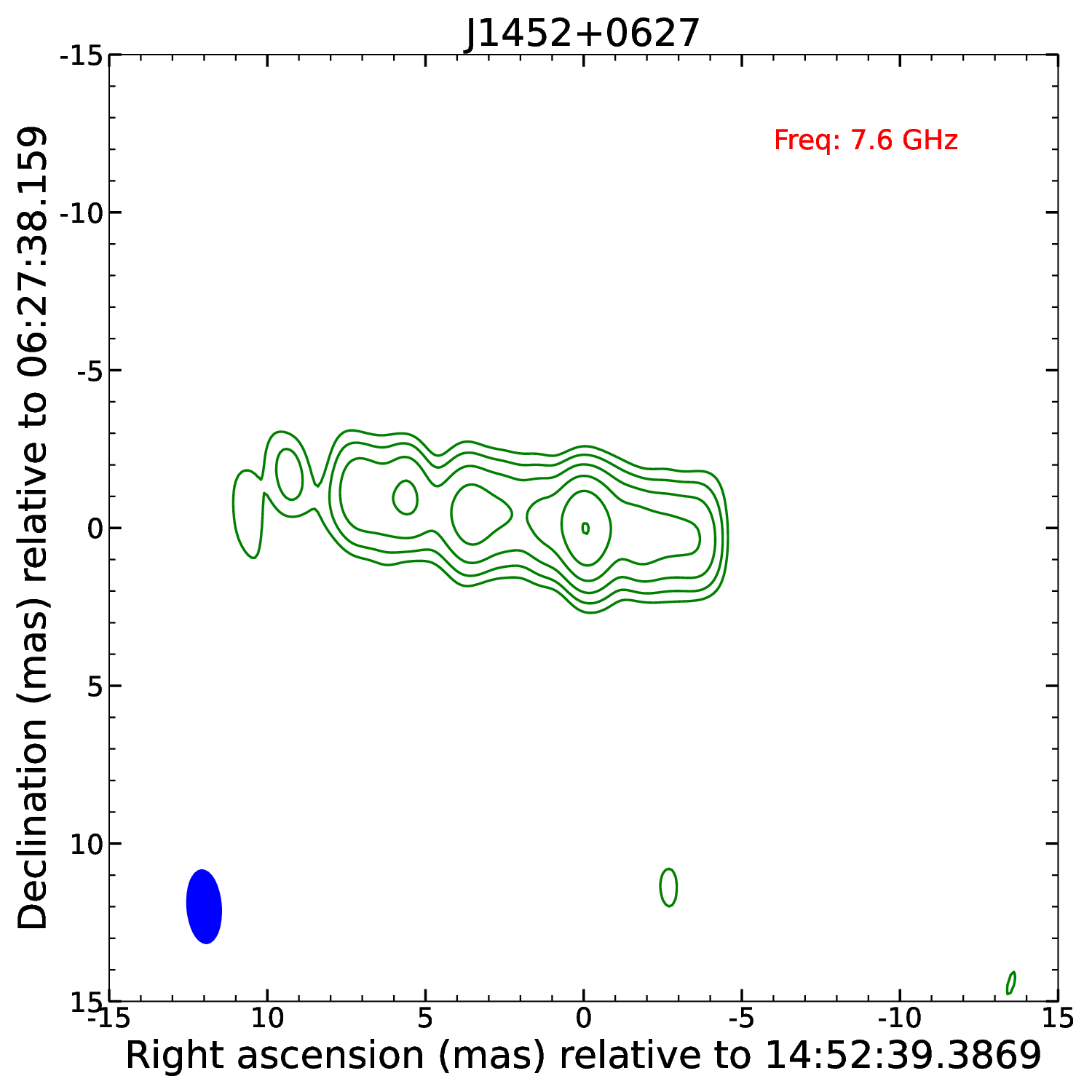}
    \caption{Left: the VLBI image of J1452+0627 at C band. The rms noise level is 0.2 mJy beam$^{-1}$ and the spatial resolution is 4.3 mas $\times$ 1.9 mas. The beam is shown in the lower-left corner. The contours are at 1.0, 2.0, 4.0, 8.0, 16.0, 32.0, 53.0 mJy beam$^{-1}$. Right: the VLBI image of J1452+0627 at X band.  The rms noise level is 0.1 mJy beam$^{-1}$ and the spatial resolution is 2.4 mas $\times$ 1.1 mas. The beam is shown in the lower-left corner. The contours are at 0.5, 1.0, 2.0, 4.0, 8.0, 16.0 mJy beam$^{-1}$. Please note that both images were taken from the Astrogeo VLBI image database.}
    \label{fig:radio_img}
\end{figure*}

SDSS J145239.38+062738.0 (hereafter J1452+0627) is a merging galaxy at redshift $z$ = 0.26716, which exhibits a tail  \citep{ahumade2020the}, see its optical image in Figure \ref{fig:J1426+0627_opt} \citep{dey2019the}. It is a type I AGN showing broad (FWHM $>2000$ km\,s$^{-1}$) $\rm H\alpha$ emission, showing high ionisation lines and is located in Seyfert region of the BPT diagram \citep[Baldwin, Philips, \& Terlevich diagram; a tool used for classifying galaxies based on emission lines;][]{baldwin1981class,murthy2021the}. In the radio band, it has been classified as a flat-spectrum quasar \citep{healey2007crate,souchay2012the}. The very long baseline interferometry (VLBI) images at C band (4.3 GHz) and X band (7.6 GHz) of J1452+2627 were obtained from the Astrogeo VLBI image database\footnote{\url{http://astrogeo.org/vlbi\_images/}} \citep{schinzel2017radio}, as shown in Figure \ref{fig:radio_img}. We note that the images were produced (by the provider) like those presented in \cite{petrov2021thewi}. The C band image has a resolution of 4.2 mas $\times$ 1.9 mas and reveals extended emission consisting of several components with a total flux of 167 mJy. The X band image has a slightly higher resolution of 2.3 mas $\times$ 1.1 mas and exhibits a similar morphology with a total flux of 66 mJy. Usually, under $\sim$ mas resolution, the radio emission from starbursts is spatially resolved. Although there is $\sim$ mas scale radio emission detected in some extreme starburst galaxies such as Mrk 273 \citep{carilli2000the,bondi2005a} and Arp 220 \citep{batejat2011resol,varenius2019the}, the morphologies are a clumpy of unresolved bean-like components that are distinct from the VLBI images of J1452+0627. More importantly, J1452+0627 is a galaxy with star formation rate of just $\sim6\,\rm M_\odot \, yr^{-1}$, see Section \ref{sec:driving_mech}, which is an order of magnitude less than those in extreme starburst galaxies. Therefore, its VLBI detected radio emission definitely come from jet. Previous observations using the Karl G. Jansky Very Large Array (JVLA) detected broad HI absorption towards it with a peak optical depth of 0.21 \citep{murthy2021the}. However, the blue side ($<-240\,\rm km\,s^{-1}$) and the red side ($>400\,\rm km\,s^{-1}$) of the HI spectrum is impacted by radio frequency interference (RFI).

In this letter, we report the detection of a fast atomic hydrogen outflow using the FAST, revealed as a shallow and blueshifted HI absorption wing that was not detected in the previous JVLA observation \citep{murthy2021the}.

\section{Observation and data reduction} \label{sec:obser}
J1452+0627 was observed in September 2022 using the ON-OFF mode as part of a programme to find 18-cm OH absorption in a sample of radio galaxies. The observations utilized the FAST L-band 19-beam receiver covering 1.00--1.50 GHz over 65,536 channels. The large bandwidth also covered the redshifted HI 21-cm line. The ON and OFF times were set to 315 seconds and 15 ON-OFF cycles were used to observe the target. During the observation, all 19 beams, each with a field of view of $\sim$3 arcmin \citep{jiang2020the}, were working simultaneously. Following \cite{zheng2020a}, we selected a specific OFF point so that when the central beam 01 was off the source, beam 14 was on the source. Therefore, both beam 01 and 14 were observing J1452+0627 and the total on-source time is 9450 seconds. The data were recorded every 0.1 seconds and the high noise diode, $\sim12.5$ K, was injected for $5\times90$ s to calibrate the flux. 

We reduced the XX and YY polarization data from beam 01 and 14 separately and then averaged them using Python-based codes written by ourselves. We first used the noise diode and the gain curves to calibrate the spectra flux. It is worth mentioning that the gain of beam 01 is different from that of beam 14, as reported in \cite{jiang2020the}. Next, for each ON-OFF cycle, the OFF-target spectrum was subtracted from the ON-target spectrum to correct the bandpass. We used two methods to subtract the continuum.

\begin{figure*}
	\includegraphics[width=18cm]{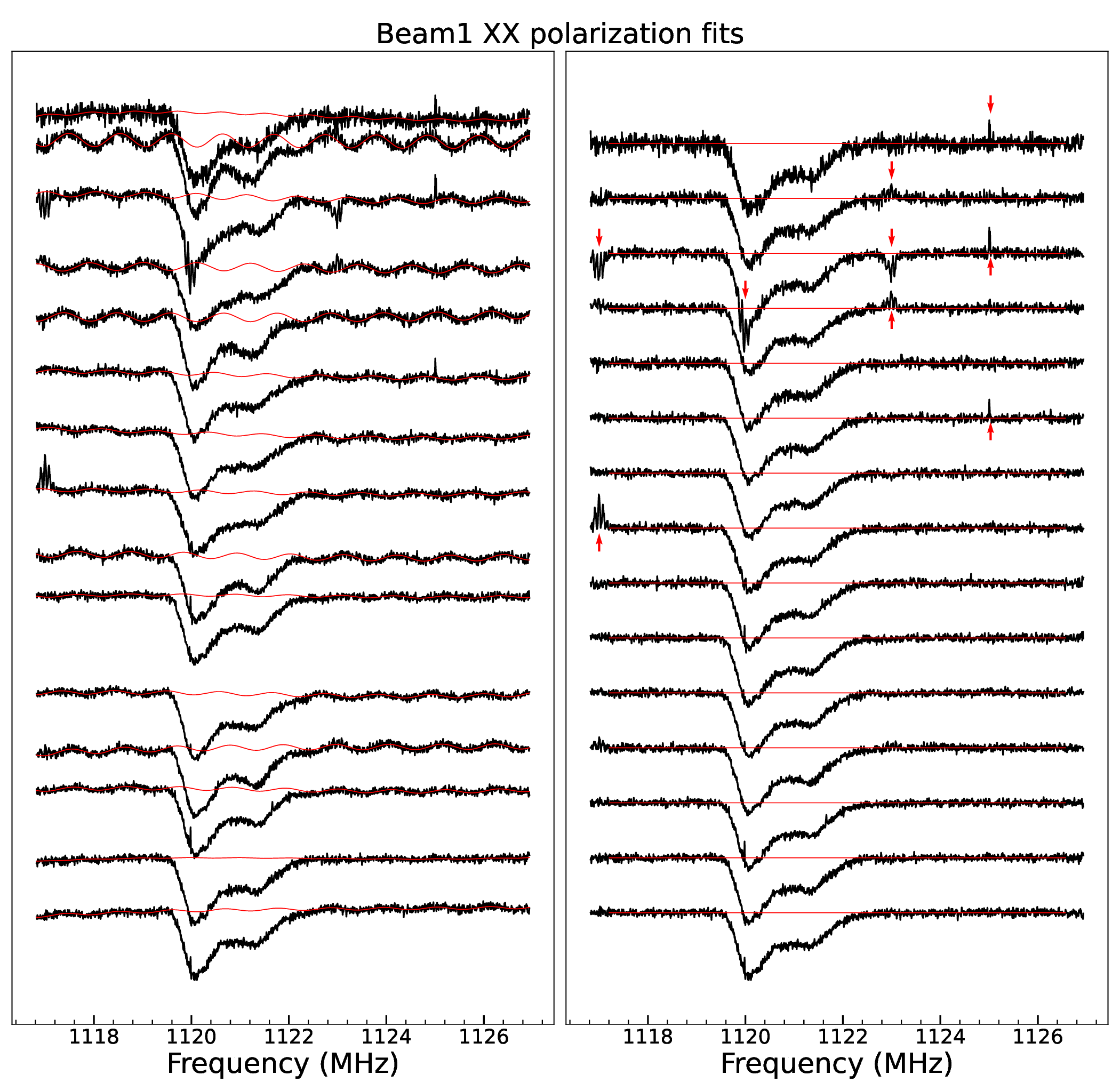}
    \caption{The continuum subtraction processes and results for XX polarization data of beam 1 using the method 1. Left: from top to bottom, the black lines are observed XX polarization spectra of beam 1 from ON-OFF cycle 1, 2, 3, ..... 13, 14, 15. The red lines show the fitted continuum. The flux scale is normalised and spectra are offset for clarity. Right: the continuum-subtracted XX polarization spectra of beam 1 from ON-OFF cycle 1, 2, 3, ..... 13, 14, 15. The red lines are horizontal. The flux scale is normalised and spectra are offset for clarity. Some RFI are indicated by red arrows. The spectral continuums look flat after continuum subtractions, indicating the continuum subtractions were good.  }
    \label{fig:beam1_polaxx_spectra_fit}
\end{figure*}

\begin{figure*}
	\includegraphics[width=18cm]{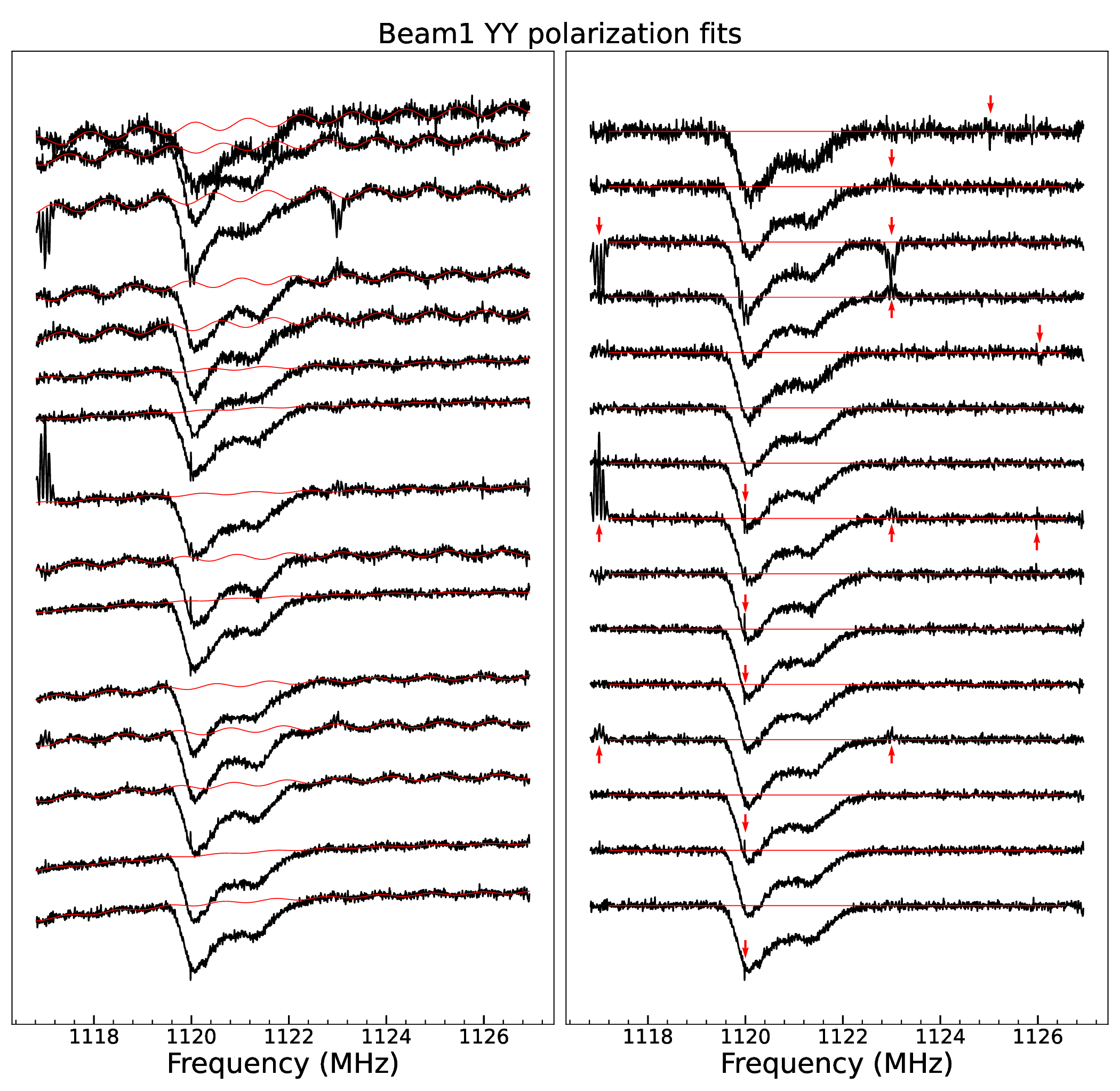}
    \caption{As Figure \ref{fig:beam1_polaxx_spectra_fit}, but for YY polarization data of beam 1.}
    \label{fig:beam1_polayy_spectra_fit}
\end{figure*}

Method 1: We implemented continuum subtraction individually for each polarization data and each ON-OFF cycle. A combination of a 3rd order polynomial and a sine functions was fitted to channels outside the range of 1119.3$\sim$1123.2 MHz to subtract the continuum. The fitting processes and results of beam 01 are shown in Figure \ref{fig:beam1_polaxx_spectra_fit}, \ref{fig:beam1_polayy_spectra_fit} while the beam 14 shows comparable results. Notably, the bandpass remained quite stable throughtout the observations and the XX and YY polarization data do not have much difference. 

To quantify the stability of the bandpass within the range of 1118$\sim$1126 MHz, we take the XX polarization data of beam 01 as an example. First, we normalised the fitted continuums of the 15 ON-OFF cycles at 1122 MHz. Then we defined the variation of the bandpass for each ON-OFF cycle as $Vari\_bp = \sqrt{\frac{\begin{matrix} \sum_{i=1}^n (\rm ContFit_{i}-ContAve_{i})^{2} \end{matrix} }{n}}$ , where $\rm ContFit_{i}$ represents the fitted continuum flux in the $i$th channel, $\rm ContAve_{i}$ is the averaged fitted continuum flux from all 15 ON-OFF cycles in the $i$th channel, and $n$ is the total number of channels within the range of 1118$\sim$1126 MHz. Finally we obtained the average value $\overline{Vari\_bp} $ for the 15 ON-OFF cycles of XX polarization data from beam 01, which is 0.0060. For YY polarization of beam 01, XX polarization of beam 14, and YY polarization of beam 14, the average value $\overline{Vari\_bp} $ were found to be 0.0066, 0.0081 and 0.0058 respectively. It is worth noting that the $\overline{Vari\_bp} $ for XX polarization of beam 14 is relatively higher, which is due to the relatively more pronounced standing waves in the first 5 ON-OFF cycles. If these cycles are excluded, the $\overline{Vari\_bp} $ would decrease to 0.0047. Although the bandpass exhibits slight variation, we do not think it would affect our detection of absorption line. This is because the spectral continuums appear flat after continuum subtractions, indicating the continuum subtractions were good and thus the influence from bandpass has been removed, see Figure \ref{fig:beam1_polaxx_spectra_fit}, \ref{fig:beam1_polayy_spectra_fit}.

\begin{figure*}
	\includegraphics[width=18cm]{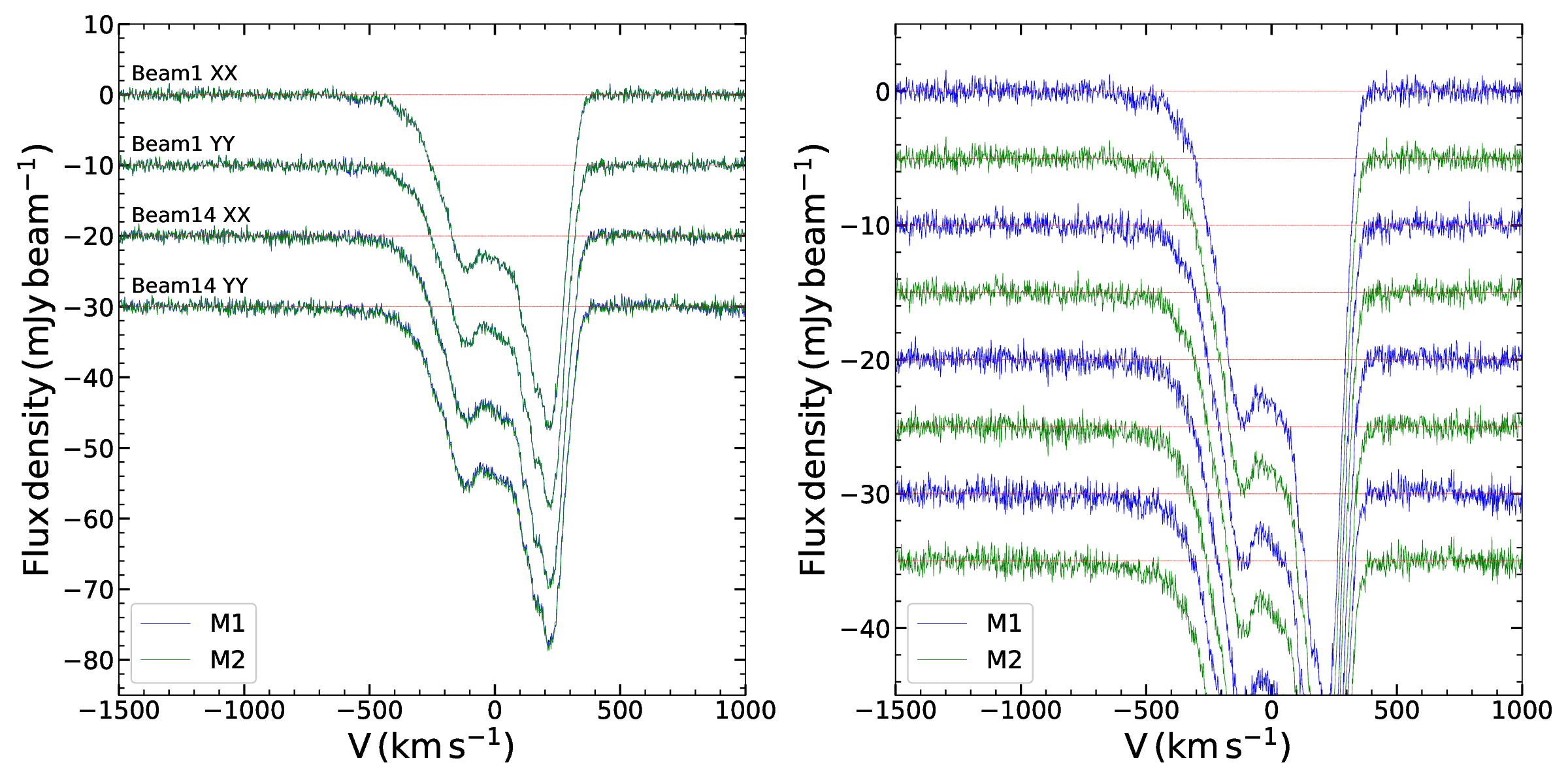}
    \caption{Left: the averaged and continuum subtracted spectra of beam 1 XX polarization, beam 1 YY polarization, beam 14 XX polarization, and beam 14 YY polarization. The beam 1 YY, beam 14 XX, and beam 14 YY are offset by -10, -20, and -30 respectively. The blue lines represent spectra produced through the method 1 while the green lines are those produced through the method 2. Right: the zoom-in plot of left but the spectra from the method 2 are more offset by -5 for a clear comparison. The method 2 gives a bit better continuum subtraction to the red side spectra from beam 14. }
    \label{fig:beam1_14_polaxxyy_spectra}
\end{figure*}

Some RFI were identified which have been indicated by red arrows. We flagged these and averaged the 15 cycles data in each polarization. The results are shown in Figure \ref{fig:beam1_14_polaxxyy_spectra}. The XX and YY polarization data from both beam 01 and beam 14 exhibit obvious blueshifted wings, providing mutual confirmation of the existence of the wing.

\begin{figure*}
	\includegraphics[width=18cm]{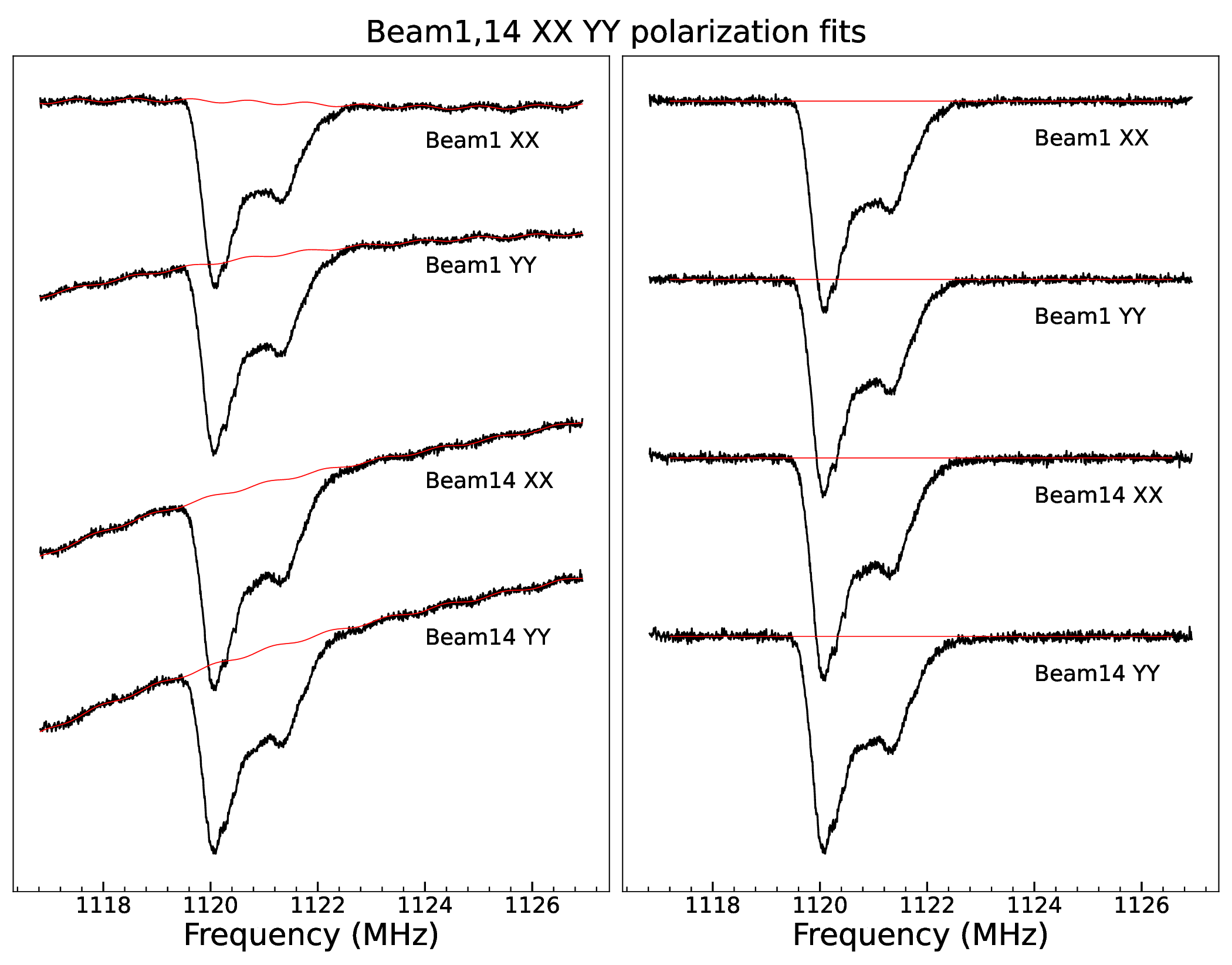}
    \caption{The continuum subtraction processes and results for XX and YY polarization data of beam 1 and 14 using the method 2. Left: from top to bottom, the black lines are spectra of beam 1 XX polarization, beam 1 YY polarization, beam 14 XX polarization, and beam 14 YY polarization, averaged from 15 ON-OFF cycles after flagging RFI. The red lines are fitted continuum. The y-axis is normalised and each next cycle spectrum is offset for clarity. Right: the black lines are corresponding continuum subtracted spectra of beam 1 XX polarization, beam 1 YY polarization, beam 14 XX polarization, and beam 14 YY polarization. The red lines are horizontal. The y-axis is normalised and each next cycle spectrum is offset for clarity. The flat continuums after continuum subtractions indicate the continuum subtractions were good as well. }
    \label{fig:beam1_14_polaxxyy_spectra_fit}
\end{figure*}

Method 2: We first flagged all the RFI and averaged the 15 ON-OFF cycles data in each polarization. Then we used a combination of a 3rd order polynomial and a sine functions to fit channels outside the range of 1119.3$\sim$1123.2 MHz to subtract the continuums. The fitting processes and results are shown in Figure \ref{fig:beam1_14_polaxxyy_spectra_fit}. The continuum subtractions were good as well. We did a comparison of the results between method 1 and method 2, which are presented in Figure \ref{fig:beam1_14_polaxxyy_spectra}. The results obtained from method 2 are remarkably similar to those acquired from method 1 except for that the red side continuum of spectra from beam 14 are flatter, indicating a bit better continuum subtractions.

Finally, the XX and YY polarization data from both beam 01 and beam 14 were averaged to generate a final spectrum. The final spectra through method 1 and method 2 were converted to the source rest frame and smoothed to 20 $\rm km\,s^{-1}$ using the Python-based code $SpectRes$ \citep{carnall2017a}, which are presented in Figure \ref{fig:J1426+0627-hi}. The final spectra exhibit clear blueshifted wings and are consistent with each other. As the method 2 provides a bit better continuum subtraction to the red side spectra from beam 14, we will use the final spectrum through the method 2 as the spectrum of J1452+0627, which has an rms noise of 0.088 mJy beam$^{-1}$. 

\begin{figure}
	\includegraphics[width=8cm]{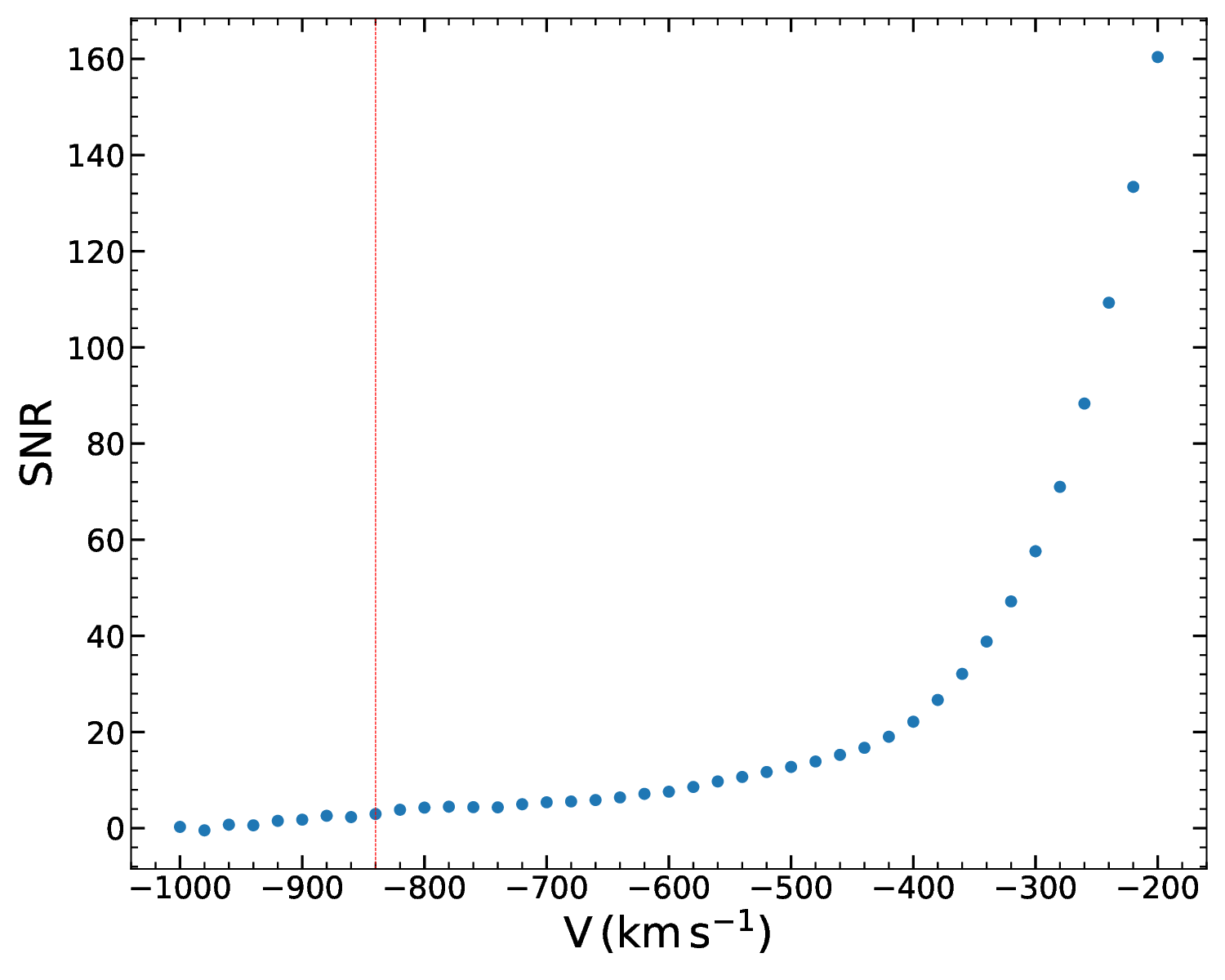}
    \caption{The SNR of the channel that smoothed from -1000 km\,s$^{-1}$ to $v$, where $-1000 \leq v \leq -200$ km\,s$^{-1}$. The vertical line marks $v$ = -840 km\,s$^{-1}$ at where the SNR = 2.95 $\simeq$ 3. We thus argue that the wing between $v$ = -1000 and -840 km\,s$^{-1}$ may be influenced by noise whereas the wing in $v > -840\, \rm km\,s^{-1}$ should be real. }
    \label{fig:snr_v}
\end{figure}

One key question is to where the blueshifted wing can extend. From Figure \ref{fig:J1426+0627-hi}, it appears that the wing extends to around -1000 km\,s$^{-1}$. However, it is crucial to consider the the influence of data processing and spectral noise. As above, we think the continuum is well subtracted and the final spectrum is primarily affected by noise. To understand the impact of noise, we smoothed the spectra from -1000 km\,s$^{-1}$ to $v$, where $-1000 \leq v \leq -200$, to one channel. The resulting channel would have a signal noise ratio (SNR) as a function of $v$, as illustrated in Figure \ref{fig:snr_v}. According to the SNR distribution, at $v$ = -840 km\,s$^{-1}$, the SNR = 2.95 $\simeq$ 3. We thus argue that the wing between $v$ = -1000 and -840 km\,s$^{-1}$ may be influenced by noise whereas the wing in $v > -840\, \rm km\,s^{-1}$ should be real.

In summary, we used two methods to do the continuum subtractions. Both methods yielded satisfactory results, providing consistent outcomes with a clear blueshifted wing, see Figure \ref{fig:J1426+0627-hi}. The outflow likely extends to -1000 km\,s$^{-1}$. Throughout the paper, the -1000 km\,s$^{-1}$ is adopted as the end point of the blueshifted wing in the calculations of its properties.

\section{Results} \label{sec:results}

\subsection{The FAST spectrum of J1452+0627}
\begin{figure*}
	\includegraphics[width=18cm]{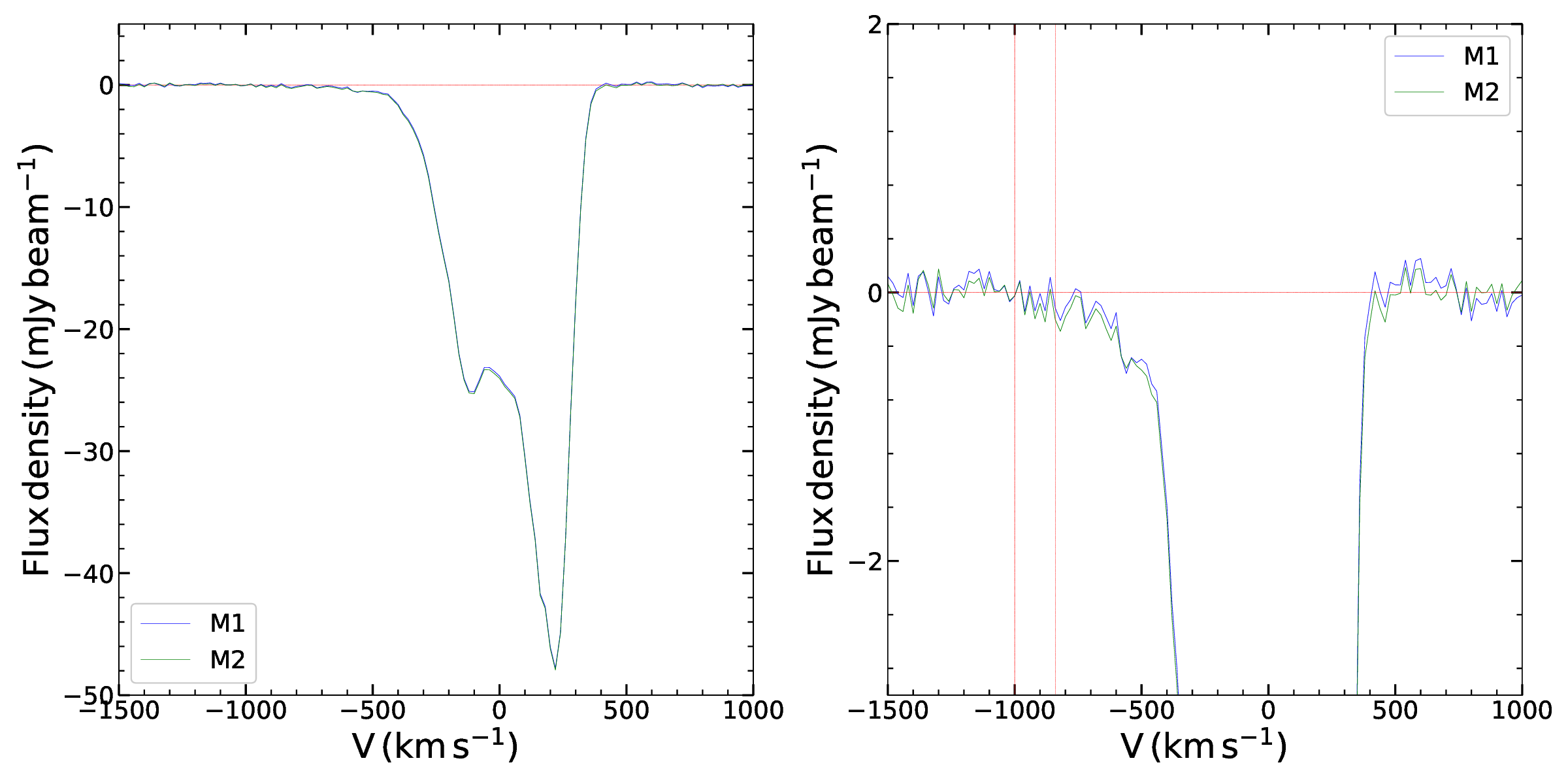}
    \caption{Left: the HI absorption spectra of J1452+0627 that shows a shallow and blueshifted wing, indicating an outflow. The blue line represents the one produced from the method 1 and the green line the one from the method 2. Both spectra are highly consistent with each other.
    Right: the zoom-in spectrum to more clearly show the shallow wing. The spectra have a velocity resolution of 20 $\rm km\,s^{-1}$. As the method 2 can give a bit better continuum subtraction to the red side spectra from beam 14, we will use the final spectrum through the method 2 as the spectrum of J1452+0627, which has an rms noise of 0.088 mJy beam$^{-1}$. The two red vertical lines mark velocities at $v$ = -1000 and -840 $\rm km\,s^{-1}$. If smoothing this velocity range to one channel, the channel would have 2.95$\sigma$ significance. }
    \label{fig:J1426+0627-hi}
\end{figure*}

\begin{figure}
	\includegraphics[width=8cm]{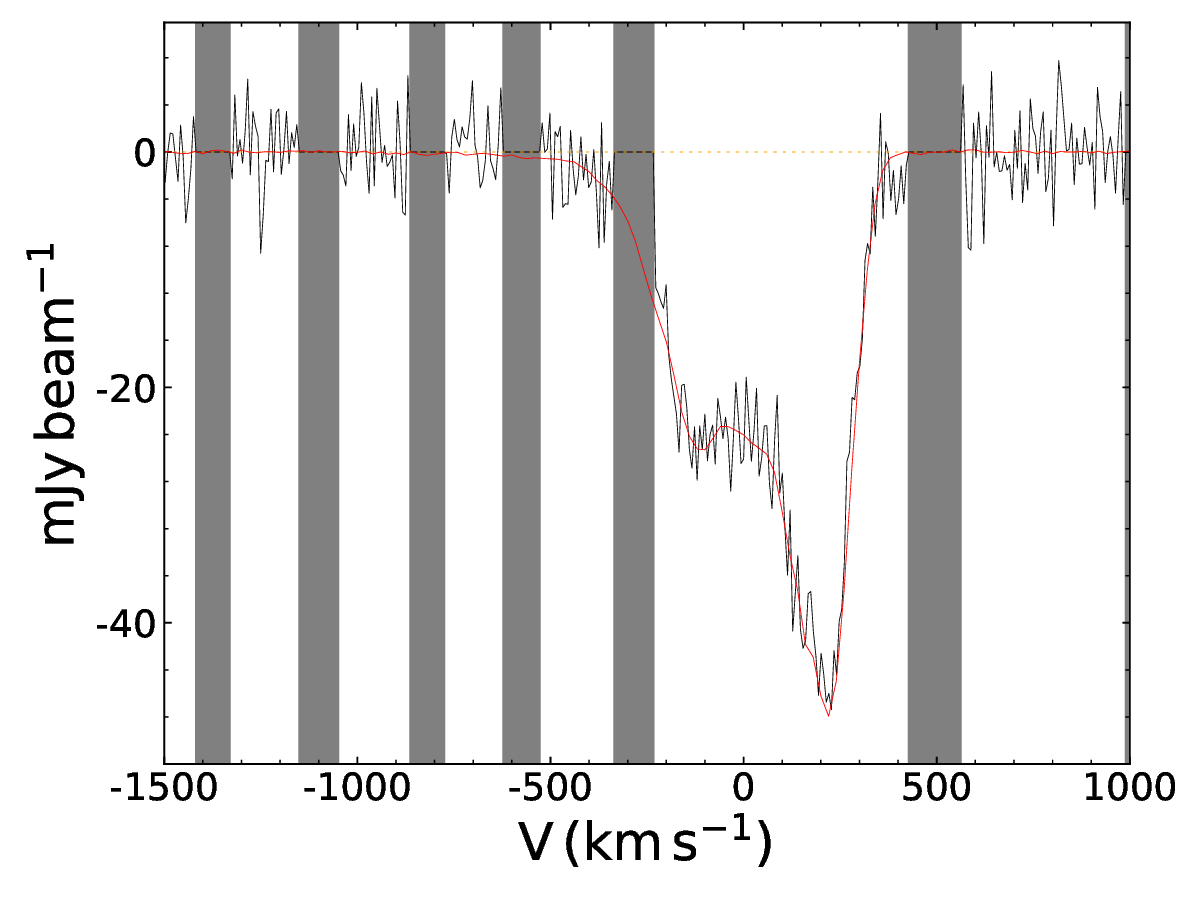}
    \caption{Comparison between our FAST spectrum (red line) and the JVLA spectrum (black line), showing consistent line parameters. The grey regions mark the channels affected by RFI in the JVLA spectrum. }
    \label{fig:FAST_JVLA_overlay}
\end{figure}

The reduced HI absorption spectrum of J1452+0627 is presented in Figure \ref{fig:J1426+0627-hi}. A shallow and blueshifted wing is evident in the spectrum. The entire spectrum is broad, spanning from $\sim-1000\,\rm km\,s^{-1}$ to $\sim500\,\rm km\,s^{-1}$. This profile aligns with the observations made by the JVLA \citep{murthy2021the}, as shown in the comparison depicted in Figure \ref{fig:FAST_JVLA_overlay}. However, our spectrum is less affected by RFI and has more than one order of magnitude higher sensitivity. Consequently our spectrum reveals a previously undetected shallow and blueshifted wing extending up to velocity of $\sim-1000\,\rm km\,s^{-1}$. 

We note that the detected absorption spectrum is a superposition of several Gaussian components arising from the absorbing gas. Since J1452+0627 is a merging galaxy, it suggests that the disk may be disturbed, thereby potentially increasing the velocity dispersion. Although the two deep absorption features in our spectrum may arise from disturbed disk gas, it is unlikely that this would lead to velocities up to $\sim -1000\,\rm km\,s^{-1}$. It is worth noting that the optical image of J1452+0627 taken by the Hubble Space Telescope (HST) reveals a dust lane aligned along the radio emission \citep{murthy2021the}, implying that the radio jet is very likely expanding into a gas structure, thereby driving an outflow. Therefore, we propose that the observed blueshifted wing traces an outflow.
 
In our observation, we measured a continuum flux density of about $302\pm4$ $\rm mJy\,beam^{-1}$, where the error was determined from the uncertainty of the gain of FAST \citep{jiang2020the}, that is consistent with the JVLA observation which gave a peak flux density of 228.4 $\rm mJy\,beam^{-1}$ and an integral flux density of 300.0 mJy. 

In \cite{murthy2021the}, the JVLA absorption spectrum was fitted with a disc model. However, the velocity dispersion fitted is quite high, $\sim45\,\rm km\,s^{-1}$ compared to a typical value of $\sim10\,\rm km\,s^{-1}$ \citep{struve2010cold}, and the fitting would be worse if a lower velocity dispersion is used \citep{murthy2021the}, which is not surprising as the galaxy is a merging galaxy and thus any disk may have been disturbed or destroyed, indicating the HI spectrum cannot be interpreted with a regularly rotating disc model. Nevertheless, it is expected that most of the HI absorption arises from the disk, including the presence of the two deep absorption features. 

To estimate the HI outflow properties, we adopted the blue wing part with $-1000<v<-500\rm\, km\,s^{-1}$ as a lower limit, and considered the entire blue side $v<0\rm\, km\,s^{-1}$ as an upper limit. Please note that these limits are reasonable. The absorption within the range of $-1000\rm\, km\,s^{-1}$ to $-500\rm\, km\,s^{-1}$ only represents a small fraction of the overall absorption, as depicted in Figure \ref{fig:J1426+0627-hi}. The absorption strength at $v=-500\,\rm km\,s^{-1}$ is approximately $\sim -0.6\,\rm mJy\,beam^{-1}$, corresponding to an optical depth of 0.002. This feature was not detected in the JVLA observation and seems cannot be accounted for in the disc model fitting \citep{murthy2021the}.  Thus, it is reasonable to consider this small portion as a lower limit. Additionally, taking the entire blue side ($v<0\,\rm km\,s^{-1}$) as an upper limit is also justified, as it likely contains a significant amount of absorption originating from the disc.

In HI absorption studies, the spin temperature $T_{\rm s}$ plays a crucial role in determining the HI column density. Usually, a fiducial value of 100 K \citep[e.g.][]{aditya2018a,sadler2020a} is often adopted when there is no direct measurement available for it. In the case of J1452+0627, considering that the VLBI images have scales of several tens of parsecs, it is reasonable to assume that the absorbing HI gas is located close to the central AGN. In such a scenario, a spin temperature of 1000 K seems plausible. Therefore, for the subsequent calculations, we will consider a range of 100 K $< T_{\rm s} < $ 1000 K.

\subsection{Outflow parameters}
To estimate the mass and energy outflow rate, we use the formulae below \citep{heckman2002galac}: 

\begin{equation}\label{equ:column_density}
{\rm d}N_{\rm HI} = 1.82\times10^{18}\,T_{\rm s}\times\tau(v) {\rm d}{v} 
\end{equation}

\begin{equation}\label{equ:partial_mass_outflow_rate}
{\rm d}\dot{M} = 30 \frac{\Omega}{4\pi} \frac{r_*}{1\,\mathrm{kpc}} \frac{{\rm d}N_{\mathrm{HI}}}{10^{21}\mathrm{cm}^{-2}} \frac{v}{300\,\mathrm{km\,s}^{-1}} \rm M_\odot\,{\rm yr}^{-1}
\end{equation}

\begin{equation}\label{equ:mass_outflow_rate}
\dot{M} = \int {\rm d}\dot{M}
\end{equation}

\begin{equation}\label{equ:energy_outflow_rate}
\dot{E} = \int \frac{1}{2} {\rm d}\dot{M} v^{2},
\end{equation}

\noindent where $T_{\rm s}$ is spin temperature, $\tau$ is optical depth and equals to $-\ln[1+\Delta S/(C_{\rm f} S_{\rm cont})]$ where $\Delta S$ is absorbed flux, $C_{\rm f}$ is covering factor, and $S_{\rm cont}$ is continuum flux, $N_{\rm HI}$ is HI column density, $\Omega$ is solid angle of outflow, $r_*$ is radius of outflow, $v$ is velocity of outflow, $\dot{M}$ is mass outflow rate, and $\dot{E}$ is energy outflow rate. While we can directly measure some parameters, we have to assume some other parameters. We assumed the covering factor is unit, which is an upper limit and often adopted when there is no direct measure it \citep[e.g.][]{aditya2018a,aditya2018agi,su2022flash}. We further assumed the $\Omega$ is $\pi$ that is often used \cite[e.g.][]{morganti2005fast,aditya2018agi} and the $r_*$ is half of the projected size of VLBI X-band emission \citep{murthy2021the}, $\sim$ 25 pc. We also considered the velocity turbulence as this would give more accurate estimates, like the case of 3C 293 \cite[e.g.][]{mahony2013the}. Finally, by integrating from $-1000\rm\,km\,s^{-1}$ to $-500\rm\,km\,s^{-1}$ and to $0\rm\,km\,s^{-1}$ and using 100 K $<T_{\rm s}<$ 1000 K, we obtained a minimum mass outflow rate of $2.8\times10^{-2}\,\rm M_\odot \, yr^{-1}$ and a maximum mass outflow rate of $3.6\,\rm M_\odot \, yr^{-1}$, corresponding to a minimum energy outflow rate of $4.2\times10^{39}\rm\,erg\,s^{-1}$ and a maximum energy outflow rate of $9.7\times10^{40}\rm\,erg\,s^{-1}$. We have summarized the outflow properties in Table \ref{tab:J1452+0627_properties}.

\begin{table}
\begin{center}
\caption{The properties of J1452+0627 }
\begin{tabular}{lc}
\hline \noalign {\smallskip}
\multicolumn{2}{c}{J1452+0627}  \\

\hline \noalign {\smallskip}

Mass outflow rate: & $2.8\times10^{-2}$--$3.6\,\,\rm M_\odot \, yr^{-1}$ \\
Energy outflow rate: & $4.2\times10^{39}$--$9.7\times10^{40}\,\,\rm erg\,s^{-1}$\\
Jet power: & $\sim5.8\times10^{43}\,\,\rm erg\,s^{-1}$ \\
SMBH mass: & $\sim1.6\times10^{8}\rm\,\,M_\odot$ \\
Broad $\rm H\alpha$ luminosity: & $\sim1.1\times10^{42}\,\,\rm erg\,s^{-1}$ \\
AGN bolometric luminosity: & $\sim2.4\times10^{44}\,\,\rm erg\,s^{-1}$\\
Broad [O III] $\lambda$5007 luminosity: & $\sim5.5\times10^{41}\,\,\rm erg\,s^{-1}$\\
Broad [O III] $\lambda$5007 FWHM: &1454 km\,$s^{-1}$\\
Narrow [O III] $\lambda$5007 luminosity: &$\sim8.5\times10^{41}\,\,\rm erg\,s^{-1}$\\
Narrow [O III] $\lambda$5007 FWHM: &690 km\,$s^{-1}$\\
$\rm[$N II] $\lambda$6584 luminosity: &$\sim8.3\times10^{41}\,\,\rm erg\,s^{-1}$\\
$\rm[$O II] $\lambda$3727 luminosity: &$\sim2.9\times10^{41}\,\,\rm erg\,s^{-1}$\\
Star formation rate: & $\sim6\,\,\rm M_\odot \, yr^{-1}$
\\

\hline \noalign {\smallskip}
\end{tabular}
\label{tab:J1452+0627_properties}
\medskip
\end{center}
\end{table}

\section{Discussion}\label{sec:disc}

\subsection{Previous observations of J1452+0627}
 
\begin{figure*}
	\includegraphics[width=18cm]{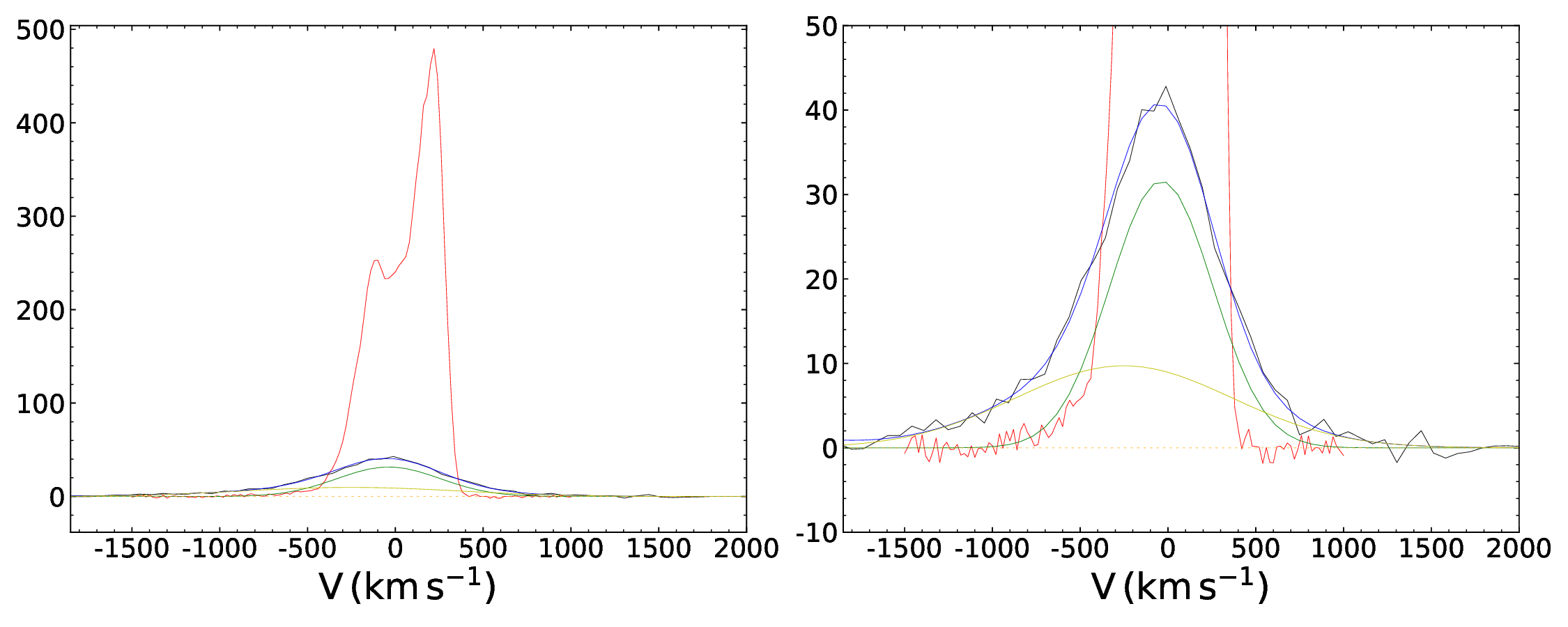}
    \caption{Left: The inverted FAST HI absorption line overlaid on the [O III] $\lambda$5007 line. The y-axis is normalised. The red line is the inverted FAST HI absorption line. The black line is the [O III] $\lambda$5007 line. The blue line is fitted [O III] $\lambda$5007 line. The green line is fitted narrow [O III] $\lambda$5007 component. The yellow line is fitted broad [O III] $\lambda$5007 component. Right: the zoom-in of the left to make a clearer comparison. }
    \label{fig:HI_OIII}
\end{figure*}

J1452+0627 was detected in HI absorption in JVLA observations and showed a full width at zero intensity (FWZI) $\sim800\,\rm km\,s^{-1}$ \citep{murthy2021the}. Besides, it has been also observed by the SDSS. We fitted the SDSS spectrum of J1452+0627 with PyQSOFit \citep{guo2018pyqso} that can simultaneously correct the Galaxy extinction, fit spectra with galaxy and AGN components, fit emission lines with Gaussian functions and drive emission line properties. Some key fitting results are summarized in Table \ref{tab:J1452+0627_properties}. The [O III] $\lambda$5007 line is best fitted with two components of which one has a FWHM $690\,\rm km\,s^{-1}$ while the another has a FWHM $1454\,\rm km\,s^{-1}$ and blueshifted by $254\,\rm km\,s^{-1}$, which is consistent with the fitting results in \cite{murthy2021the}. In the JVLA observations where the HI absorption width is close to that of the narrow [O III] $\lambda$5007 component, \cite{murthy2021the} suggested these two phases of the gas might be associated. Our FAST observation that has much lower rms noise revealed that the HI line could be as broad as $\sim1500\,\rm km\,s^{-1}$ (FWZI) with a shallow component blueshifted up to $\sim-1000\,\rm km\,s^{-1}$, suggesting that we maybe detected the neutral counterpart of the broad and blueshifted optical kinematic component. We have overlaid the HI absorption profile on the [O III] $\lambda$5007 line profile in Figure \ref{fig:HI_OIII}.

\subsection{Driving mechanism}\label{sec:driving_mech}
 
A key question is what drives the outflow. Both starburst and AGN can produce outflows \citep[e.g.][]{di2005energ,veilleux2005galac,booth2009cosmo,morganti2018the,veilleux2020cool}. In the case of J1452+0627, which is identified as an AGN, estimating the star formation rate through the commonly used $\rm H\alpha$ line is not applicable. However, an alternative method based on the recalibration of the [O II]$\lambda$3727 emission line can be employed to estimate the star formation rate in AGN \citep{zhuang2019recal}. By fitting the SDSS spectrum of J1452+0627, we obtained its [N II] $\lambda$6584, [O II] $\lambda$3727 and narrow [O III] $\lambda$5007 emission lines luminosity and then derived a star formation rate of $\sim6\,\rm M_\odot \, yr^{-1}$ using the Equation 7 in \cite{zhuang2019recal}.

The mass outflow rate that can be driven by this star formation rate was estimated as $\sim1.5\,\rm M_\odot \, yr^{-1}$ following Equation 1 in \cite{veilleux2005galac}, which is larger than or consistent with the observed mass outflow rate in J1452+0627, indicating that the star formation is able to drive the observed outflow.

The AGN could be the reasonable power source of the HI outflow. We estimated the AGN bolometric luminosity using $L_{\rm bol}\simeq30L_{\rm BLR}$ \citep{xu2009on}, where $L_{\rm BLR}$ is the luminosity of broad-line region and can be estimated from broad $\rm H\alpha$ line \citep{gelotti1997jets}. By fitting the spectrum, we found the $\rm H\alpha$ line can be fitted with a broad component (FWHM $>2000$\,km\,s$^{-1}$). The broad component has a luminosity of $\sim1.1\times10^{42}\,\rm erg\,s^{-1}$ which gives a bolometric luminosity of $\sim2.4\times10^{44}\,\rm erg\,s^{-1}$. 

Therefore, the ratio between energy outflow rate and AGN bolometric luminosity is between $1.7\times10^{-5}$ and $4.0\times10^{-4}$ which is much smaller than those in AGN energy-conserving wind-driven outflows \citep[e.g.][]{king2010black,zubovas2012clear,costa2014feedb} and those required to produce the $M-\sigma$ relation and powerful feedback \citep[e.g.][]{di2005energ,booth2009cosmo,hopkins2010quasa}, indicating the AGN radiation has enough power to drive the HI outflow.

Simulations and observations suggest that radio jets can drive powerful outflows when they expand through the ISM of the host galaxy \citep[e.g.][]{wagner2011relat,wagner2012drivi,mahony2013the,morganti2013radio}. A radio jet could drive gas to high velocity if the ratio of jet power to Eddington luminosity $\eta = P_{\rm jet}/L_{\rm Edd}$ is above $10^{-4}$ \citep[e.g.][]{wagner2011relat,wagner2012drivi}. To estimate the jet power, it is necessary to determine the radio emission contribution from star formation first. Given a star formation rate of $\sim6\,\rm M_\odot \, yr^{-1}$, the Equation 3 in \cite{sullivan2001a} provides an estimate of the maximum luminosity of $5.3\times10^{28}\rm\,erg\,s^{-1}\,Hz^{-1}$ at 1.4 GHz. However, with FAST observations,  a continuum flux density of $302\pm4$ $\rm mJy$ at 1.4 GHz was obtained, corresponding a luminosity of $6.7\times10^{32}\rm\,erg\,s^{-1}\,Hz^{-1}$. This indicates the radio contribution from star formation is negligible. By using the Equation 2 in \cite{best2006agn}, the jet power is estimated to be $P_{\rm jet}\sim5.8\times10^{43}\,\rm erg\,s^{-1}$.

The SMBH mass can be estimated from the M-$\sigma$ relation \citep{mcconnell2013revis}. We used the Penalized PiXel-Fitting (pPXF) method \citep{cappellari2004param,cappellari2017impro,cappellari2022full} to fit the SDSS spectrum and obtained a stellar velocity dispersion $190\pm25.6\,\rm km\,s^{-1}$ which gives a SMBH mass $\sim1.6\times10^{8}\,\rm M_\odot$. Therefore, the $\eta$ in J1452+0627 is $\sim2.8\times10^{-3}$, meaning the radio jet is able to drive the observed HI outflow.

In summary, it seems the star formation, the AGN radiation, and the jet all can alone drive the observed HI outflow. Here we try to have a deeper analysis based on the HI outflow kinematics. The HI spectrum has a velocity likely blueshifted up to $\sim1000\,\rm km\,s^{-1}$ and a broad velocity width with a FWZI possibly up to $\sim1500\,\rm km\,s^{-1}$. Optical Na I D absorption line is a good indicator of HI gas. Previous Na I D absorption line observations have revealed that the HI outflows produced by large starbursts have an average FWHM of $\sim 275 \rm km\,s^{-1}$ and an average maximum outflow velocity $\sim 300-400 \rm km\,s^{-1}$ \citep{veilleux2005galac}. In comparison, it is very unlikely that the HI outflow in J1452+0627 is driven by starbursts. Broad Na I D absorption lines, typically $>$500 -- 600 km\,s$^{-1}$, have been detected before, but in jet-driven outflows \citep{lehnert2011the}. AGN radiation-driven outflows can have ionized gas with FWHM $\sim$ a few thousands km\,s$^{-1}$ and blueshifted by up to a few ten thousands km\,s$^{-1}$, like those in broad absorption line (BAL) quasars \citep{weymann1991compa,gibson2009a}. However, it is not clear if the neutral gas in radiation-driven outflows can have similar kinematics or resemble that in our detected HI outflow. As of now, there is no observations to support this. In contrast, the kinematics of the HI outflow in J1452+0627 resemble those in jet-gas interactions \citep[e.g.][]{morganti2005fast,lehnert2011the}. For example, the jet-driven HI outflow in 4C\,12.50 has a FWZI of $\sim1600\rm km\,s^{-1}$ and a maximum outflow velocity of $\sim1200\rm km\,s^{-1}$ \citep{morganti2005fast,morganti2013radio}. Similarly, in 3C 293, the jet-driven HI outflow has a FWZI of $\sim1400\rm km\,s^{-1}$, of which $\sim1000\rm km\,s^{-1}$ are blueshifted \citep{mahony2013the}. More cases of jet-driven HI outflows that have similar kinematics can be seen in \cite{morganti2005fast,morganti2018the}. Therefore, based on the similar kinematics to know jet-driven outflows, the radio jet has the relatively higher probability, compared to the AGN radiation and star burst, to be the driver of the HI outflow in J1452+0627. But, we suggest that VLBI spectral line observations can help us locate the position of the outflowing gas and thus pin down whether it is the radio jet that drives the outflow as the cases studied in, for example, 4C\,12.50 \citep{morganti2013radio} and 3C\,293 \citep{mahony2013the}.

\subsection{Impact on the host galaxy}
Based on whether the star formation is enhanced or suppressed, AGN-driven feedback is classified as `positive' or `negative'. In this section, we will discuss whether the power source of the HI outflow in J1452+0627 can enhance or suppress star formation in the host galaxy.  In hydrodynamical simulations setting $\sim5$ per cent of AGN radiation can be injected to surrounding gas, it was found that the feedback would be powerful enough to quench star formation \citep{di2005energ}. However, in a `two-phase' model, an order of magnitude less injecting energy, $\sim0.5$ per cent of AGN radiation, is capable of driving outflows to destroy cold gas reservoir and/or clear the gas out of the galaxy \citep{hopkins2010quasa}. The ratio between the energy outflow rate and the AGN bolometric luminosity in J1452+0627 is between $1.7\times10^{-5}$ and $4.0\times10^{-4}$. Therefore, if the HI outflow is driven by AGN radiation, the AGN radiation seems not powerful enough to provide negative feedback.

As discussed in the last section, it is more likely the HI outflow is driven by the radio jet which has a power of $\sim5.8\times10^{43}\,\rm erg\,s^{-1}$. In hydrodynamical simulations that have pc scale spatial resolution to study the interactions between jets and ISM \citep{wagner2011relat,wagner2012drivi}, it was found that the jet-induced feedback is effective enough to disperse cloudy gas and thus inhibit star formation if jets have power in the range of $10^{43}$--$10^{46}$ erg\,s$^{-1}$ and jets power to Eddington luminosity $\eta = P_{\rm jet}/L_{\rm Edd}$ is above $10^{-4}$. In J1452+0627, the $P_{\rm jet}\sim5.8\times10^{43}\,\rm erg\,s^{-1}$ and the $\eta\sim 2.8\times10^{-3}$, therefore the jet seems have the potential to provide negative feedback.

In summary, if the HI outflow is driven by the AGN radiation, the AGN radiation seems not powerful enough to provide negative feedback whereas the radio jet seems have the potential to provide negative feedback.

\section{Conclusion}\label{sec:conlcusion}

In this paper, we report the discovery of a fast neutral outflow in J1452+0627 with the FAST. The outflow exhibits a blueshifted wing extending up to -1000 km\,s$^{-1}$. The absorption strength of the outflow at -500 km\,s$^{-1}$ is approximately $-0.6\,\rm mJy\,beam^{-1}$, corresponding to an optical depth of 0.002. The weak absorption strength and low optical depth indicate that detecting such outflows is very challenging because it is hard to go down to such low optical depth towards most radio sources within reasonable on-source times by using most current radio telescopes. The FAST telescope with its large collecting area is a fantastic tool to detect such faint outflows as demonstrated by our observations. Our discovery represents the first detection of a fast outflow by the FAST. It is expected more such detections will be made with the FAST telescope. So far, the detections of fast, broad, and shallow neutral outflows are rare, see the compilation by \cite{morganti2018the}.

The main results associated with the outflow are summarized below:

$\bullet$ The outflow exhibits  a mass outflow rate ranging from $2.8\times10^{-2}\,\rm M_\odot \, yr^{-1}$ to $3.6\,\rm M_\odot \, yr^{-1}$, corresponding to an energy outflow rate between $4.2\times10^{39}\,\rm erg \,s^{-1}$ and $9.7\times10^{40}\,\rm erg \,s^{-1}$, assuming 100 K $<T_{\rm s}<$ 1000 K.

$\bullet$ It seems the outflow is more likely driven by the jet based on the kinematics, although the starbursts and the AGN radiation cannot be ruled out as power sources. To elucidate the driving mechanism of the outflow and more accurately estimate the impact on the host galaxy, VLBI spectral observations are needed to localise the outflow.

$\bullet$ The ratio between energy outflow rate and AGN bolometric luminosity was found to be between $1.7\times10^{-5}$ and $4.0\times10^{-4}$, which is smaller than that required to strikingly impact gas reservoir and thus suppress star formation \citep{hopkins2010quasa}, indicating that if the HI outflow is driven by the AGN radiation, the AGN radiation seems not powerful enough to provide negative feedback.

$\bullet$ The jet in J1452+0627 seems have the potential to provide negative feedback because the jet has a power of $P_{\rm jet}\sim5.8\times10^{43}\,\rm erg\,s^{-1}$ and the ratio between jet power and Eddington luminosity is $2.8\times10^{-3}$ which meet the criteria to inhibit star formation in simulations \citep{wagner2011relat,wagner2012drivi}.

\begin{acknowledgments}
We thank the anonymous referee for the constructive comments which have greatly improved the draft.

We thank the staff of the FAST who have made these observations possible. We thank Suma Murthy who provided the JVLA spectrum of J1452+0627 and Hengxiao Guo who fitted the SDSS spectrum of J1452+0627 with the pPXF.

This work made use of the data from FAST (Five-hundred-meter Aperture Spherical radio Telescope).  FAST is a Chinese national mega-science facility, operated by National Astronomical Observatories, Chinese Academy of Sciences.

This work made use of the Astrogeo VLBI FITS image database.

The Legacy Surveys consist of three individual and complementary projects: the Dark Energy Camera Legacy Survey (DECaLS; Proposal ID \#2014B-0404; PIs: David Schlegel and Arjun Dey), the Beijing-Arizona Sky Survey (BASS; NOAO Prop. ID \#2015A-0801; PIs: Zhou Xu and Xiaohui Fan), and the Mayall z-band Legacy Survey (MzLS; Prop. ID \#2016A-0453; PI: Arjun Dey). DECaLS, BASS and MzLS together include data obtained, respectively, at the Blanco telescope, Cerro Tololo Inter-American Observatory, NSF’s NOIRLab; the Bok telescope, Steward Observatory, University of Arizona; and the Mayall telescope, Kitt Peak National Observatory, NOIRLab. Pipeline processing and analyses of the data were supported by NOIRLab and the Lawrence Berkeley National Laboratory (LBNL). The Legacy Surveys project is honored to be permitted to conduct astronomical research on Iolkam Du’ag (Kitt Peak), a mountain with particular significance to the Tohono O’odham Nation.

NOIRLab is operated by the Association of Universities for Research in Astronomy (AURA) under a cooperative agreement with the National Science Foundation. LBNL is managed by the Regents of the University of California under contract to the U.S. Department of Energy.

This project used data obtained with the Dark Energy Camera (DECam), which was constructed by the Dark Energy Survey (DES) collaboration. Funding for the DES Projects has been provided by the U.S. Department of Energy, the U.S. National Science Foundation, the Ministry of Science and Education of Spain, the Science and Technology Facilities Council of the United Kingdom, the Higher Education Funding Council for England, the National Center for Supercomputing Applications at the University of Illinois at Urbana-Champaign, the Kavli Institute of Cosmological Physics at the University of Chicago, Center for Cosmology and Astro-Particle Physics at the Ohio State University, the Mitchell Institute for Fundamental Physics and Astronomy at Texas A\&M University, Financiadora de Estudos e Projetos, Fundacao Carlos Chagas Filho de Amparo, Financiadora de Estudos e Projetos, Fundacao Carlos Chagas Filho de Amparo a Pesquisa do Estado do Rio de Janeiro, Conselho Nacional de Desenvolvimento Cientifico e Tecnologico and the Ministerio da Ciencia, Tecnologia e Inovacao, the Deutsche Forschungsgemeinschaft and the Collaborating Institutions in the Dark Energy Survey. The Collaborating Institutions are Argonne National Laboratory, the University of California at Santa Cruz, the University of Cambridge, Centro de Investigaciones Energeticas, Medioambientales y Tecnologicas-Madrid, the University of Chicago, University College London, the DES-Brazil Consortium, the University of Edinburgh, the Eidgenossische Technische Hochschule (ETH) Zurich, Fermi National Accelerator Laboratory, the University of Illinois at Urbana-Champaign, the Institut de Ciencies de l’Espai (IEEC/CSIC), the Institut de Fisica d’Altes Energies, Lawrence Berkeley National Laboratory, the Ludwig Maximilians Universitat Munchen and the associated Excellence Cluster Universe, the University of Michigan, NSF’s NOIRLab, the University of Nottingham, the Ohio State University, the University of Pennsylvania, the University of Portsmouth, SLAC National Accelerator Laboratory, Stanford University, the University of Sussex, and Texas A\&M University.

BASS is a key project of the Telescope Access Program (TAP), which has been funded by the National Astronomical Observatories of China, the Chinese Academy of Sciences (the Strategic Priority Research Program “The Emergence of Cosmological Structures” Grant \# XDB09000000), and the Special Fund for Astronomy from the Ministry of Finance. The BASS is also supported by the External Cooperation Program of Chinese Academy of Sciences (Grant \# 114A11KYSB20160057), and Chinese National Natural Science Foundation (Grant \# 12120101003, \# 11433005).

The Legacy Survey team makes use of data products from the Near-Earth Object Wide-field Infrared Survey Explorer (NEOWISE), which is a project of the Jet Propulsion Laboratory/California Institute of Technology. NEOWISE is funded by the National Aeronautics and Space Administration.

The Legacy Surveys imaging of the DESI footprint is supported by the Director, Office of Science, Office of High Energy Physics of the U.S. Department of Energy under Contract No. DE-AC02-05CH1123, by the National Energy Research Scientific Computing Center, a DOE Office of Science User Facility under the same contract; and by the U.S. National Science Foundation, Division of Astronomical Sciences under Contract No. AST-0950945 to NOAO.

RZS is supported by NSFC grant No. 11988101, the National Science Foundation of China (grant 11873073), Shanghai Pilot Program for Basic Research-Chinese Academy of Science, Shanghai Branch (JCYJ-SHFY-2021-013), the National SKA Program of China (Grant No. 2022SKA0120102), the science research grants from the China Manned Space Project with NO. CMSCSST-2021-A06, and the Original Innovation Program of the Chinese Academy of Sciences (E085021002).

MFG is supported by the National Science Foundation of China (grant 11873073), Shanghai Pilot Program for Basic Research-Chinese Academy of Science, Shanghai Branch (JCYJ-SHFY-2021-013), the National SKA Program of China (Grant No. 2022SKA0120102), the science research grants from the China Manned Space Project with NO. CMSCSST-2021-A06, and the Original Innovation Program of the Chinese Academy of Sciences (E085021002).

DL is supported by NSFC grant No. 11988101.

ZZ is supported by NSFC grant No. 11988101, 12041302, and U1931110. ZZ is also supported by the science research grant from the China Manned Space Project with grant no. CMS-CSST-2021-A08.

N.-Y. Tang is sponsored by Zhejiang Lab Open Research Project (NO. K2022PE0AB01), Cultivation Project for FAST Scientific Payoff and Research Achievement of CAMS-CAS, National key R\&D program of China under grant No. 2018YFE0202900 and the University Annual Scientific Research Plan of Anhui Province (NO.2022AH010013).

\end{acknowledgments}

%





\bibliography{J1452+0627}{}
\bibliographystyle{aasjournal}



\end{document}